\documentclass[aps,twocolumn,showpacs,preprintnumbers,amsmath,amssymb,nofootinbib,superscriptaddress,showkeys]{revtex4}

\usepackage{epsfig}
\usepackage{graphicx}

\begin{document}

\title{Renormalization of the Off-shell chiral two-pion exchange NN
  interactions~\footnote{Supported by Ministerio de Ciencia y
  Tecnolog\'\i a under Contracts No. FPA2007-65748, 
  by Junta de
  Castilla y Le\'on under Contract No.  SA-106A07 and GR12, by the
  European Community-Research Infrastructure Integrating Activity
  ``Study of Strongly Interacting Matter'' (HadronPhysics2 Grant
  no. 227431), the Spanish Ingenio-Consolider 2010 Program CPAN
  (CSD2007-00042), the Spanish DGI and FEDER funds with grant
no. FIS2008-01143/FIS, Junta de Andaluc{\'\i}a grants no.  FQM225-05,
EU Integrated Infrastructure Initiative Hadron Physics Project
contract no. RII3-CT-2004-506078}}
\author{D. R. Entem}\email{entem@usal.es}
  \affiliation{Grupo de F\'{\i}sica Nuclear, IUFFyM, \\ Universidad de
  Salamanca, E-37008 Salamanca,Spain} 
\author{E. Ruiz
  Arriola}\email{earriola@ugr.es} \affiliation{Departamento de
  F\'{\i}sica At\'omica, Molecular y Nuclear, \\ Universidad de Granada,
  E-18071 Granada, Spain.}
  \date{\today}

\begin{abstract} 
\rule{0ex}{3ex} The renormalization, finiteness and off-shellness of
short distance power like singular interactions is discussed. We show
analytically that the renormalizability of the off-shell scattering
amplitude relies completely on the corresponding on-shell amplitude
without proliferation of new counterterms. We illustrate the result by
complementary calculations both in coordinate as well as in momentum
space in the simplest $^1S_0$ channel for chiral np interactions
including Two Pion Exchange.
\end{abstract}


\maketitle




A traditional source of theoretical uncertainty in the study of
nuclear physics and nuclear reactions has been the relevance and
significance of off-shellness in the NN force (For a review up to the
mid seventies see e.g. \cite{Srivastava:1975eg} and references
therein). As is well known, any off-shell ambiguity should cancel in
the final results and off-shellness itself cannot be measured as a
matter of principle. This does not necessarily mean, however, that
off-shellness can generally be completely disposed of and most few-
and many body calculations do involve off-shell quantities as
intermediate stages. This feature relies on the fundamental fact that
quantum mechanics is naturally formulated in terms of wave functions
while they are not directly measurable quantities except at
asymptotically large distances. A paralell statement for Quantum Field
Theory applies for the fields themselves as well as the associated
Green functions.

Potential approaches to the NN interaction need the half off-shell
extrapolated potential and the half off-shell T-matrix is used to
determine the on-shell $S-$matrix. Moreover, a knowledge of the
off-shell T-matrix is needed e.g. for nucleon-nucleon bremsstrahlung,
the three nucleon problem as well as nuclear matter calculations and
thus a phenomenological determination of the off-shell T-matrix has
been the subject of intense research in the
past~\cite{Srivastava:1975eg}. A relevant issue in this regard is that
the definition of off-shellness is largely conventional. Actually, the
quantum mechanical trading between two body off-shellness and three
and many body forces was shown in Ref.~\cite{1990FBS.....9...97P}, and
further discussed for potential models~\cite{Amghar:1995av} and within
Lagrangean field theory~\cite{Furnstahl:2000we}. Moreover, unitarity
for the three body problem rests on off-shell unitarity for the two
body problem, imposing constraints on the acceptable
off-shellness~\cite{Kowalski:1966zz,Kowalski:1975zz}.

Within the Effective Field Theory (EFT) approach to nuclear physics
based on chiral symmetry~\cite{Weinberg:1990rz,Ordonez:1995rz} (for
comprehensive reviews see
e.g. Ref.~\cite{Bedaque:2002mn,Epelbaum:2005pn,Machleidt:2005uz}) the
ambiguities related to off-shellness can be rephrased in the freedom
to undertake field dependent transformations and using the equations
of motion. Actually, in purely contact EFT's, where the interaction is
represented by a polynomial in momenta and/or energy, off-shellness
can be completely ignored from the start by using local field
redefinitions~\cite{Georgi:1991ch,Furnstahl:2000we}. This fact does
not generally hold for finite range interactions stemming from
particle exchange if the exchanged momentum becomes comparable to the
exchanged particle mass and where non-local and singular field
redefinitions would be needed.  Similarly to phenomenological
potentials, Chiral potentials are not free from these ambiguities
since by construction they are extracted from the S-matrix with a
given perturbative definition and it is possible in particular to
choose either energy~\cite{Ordonez:1995rz} or
momentum~\cite{Kaiser:1997mw} dependent forms by using the on-shell
condition (see also Ref.~\cite{Entem:2002sf}). Further ambiguities and
their equivalence have been discussed in Ref.~\cite{Friar:1999sj}.
Quite generally Chiral potentials are based on an expansion in inverse
powers of $f_\pi$ (the pion weak decay constant) and $M_N$ (the
nucleon mass) and are necessarily singular at short distances by
purely dimensional reasons,
\begin{eqnarray}
V(r) \to \frac{1}{M_N^m f_\pi^n r^{n+m+1}}  \, . 
\label{eq:chiral-sing}
\end{eqnarray} 
The problem on how these singularities should be handled from a
renormalization point of view has been addressed in a series of works
for the {\it on-shell} case where the finiteness can be established
{\it a
  priori}~\cite{PavonValderrama:2004td,PavonValderrama:2007nu}. In the
subtractive renormalization method conducted in momentum space
\cite{Frederico:1999ps,Timoteo:2005ia, Yang:2007hb} it has been shown
numerically that off-shell amplitudes are finite. Still, an analytic
proof would be timely since there exist examples where a direct
calculation of a Green functions in effective field theory does not
necessarily guarantee off-shell finiteness from on shell
renormalization conditions (see e.g. Ref.~\cite{Appelquist:1980ae}) and
suitable field redefinitions may be requested to ensure off-shell
renormalizability.

The hard core problem was the first singular potential which was
treated by van Leeuwen and Reiner in the early
sixties~\cite{1961Phy....27...99V}. It was shown that a finite and
smooth result for the off-shell scattering amplitude could be achieved
if the hard core boundary condition was also fulfilled by the
off-shell wave functions. Let us note that for a non-singular
potential the standard way of going off-shell is to keep the same {\it
  regular} boundary condition as in the on-shell case. In the present
paper we exploit this idea of a common boundary condition both for the
on-shell and off-shell states to show that finiteness rests on pure
on-shell properties. Moreover, we check the off-shell equivalence
between the boundary condition renormalization in coordinate and the
counterterm renormalization in momentum space thus extending similar
findings for the on-shell case~\cite{Entem:2007jg,Valderrama:2007ja}.


In the CM frame, where the np kinetic energy is given by $E = p^2 / M
$, with $M = 2 \mu_{np} = 2 M_n M_p /(M_p + M_n) $, the scattering
process is described by using the Lippmann-Schwinger equation
\begin{eqnarray}
T(E) = V + V G_0 T(E)  \, , 
\end{eqnarray}
with $V$ the potential operator and $G_0 = (E-H_0 )^{-1} $ the
resolvent of the free Hamiltonian and $T(E)$ the T-matrix for energy
$E$.  The outgoing boundary condition corresponds to $ E \to E + {\rm
  i} 0^+ $. Using the normalization $ \langle \vec x | \vec k \rangle
= e^{{\rm i} \vec k \cdot \vec x} /(2\pi)^{3/2} $ one has
\begin{eqnarray}
\langle \vec k' | T(E) | \vec k \rangle = \langle \vec k' | V | \vec k
\rangle + \int^\Lambda d^3 q { \langle \vec k' | V | \vec q \rangle
\langle \vec q | T(E) | \vec k \rangle \over E - (q^2 / 2 \mu ) } \, . 
\nonumber \\
\end{eqnarray} 
Here $\Lambda$ means a generic regulator and represents the scale
below which all physical effects are taken into account {\it
  explicitly}. Here we assume that $|\vec k \rangle$ and $|\vec k'
\rangle$ are plane-wave states with energies, $E_k= k^2/(2\mu)$,
$E_{k'}=k'^2/(2\mu)$ different from $E_p= p^2/(2\mu)$.

In coordinate space this corresponds to solve the inhomogeneous
Schr\"odinger equation,
\begin{eqnarray}
-\frac1 M \nabla^2 \Psi ( \vec x) + V (\vec x) \Psi (\vec x) &=& E_p
 \Psi (\vec x) \nonumber \\ 
&+& (E_{k}-E_{p} )e^{i \vec k' \cdot \vec x}\, , 
\label{eq:sch-coor}
\end{eqnarray} 
with the outgoing boundary condition
\begin{eqnarray}
\Psi (\vec x) \to \left[ e^{i \vec k \cdot \vec x} + f_p (\vec k', \vec 
k) \frac{e^{i p r}}{r} \right] \chi_{t,m_t}^{s,m_s} \, , 
\label{eq:bc-coor}
\end{eqnarray} 
with $f(\hat k', \hat k)$ the quantum mechanical off-shell scattering
matrix amplitude and $\chi_{t,m_t}^{s,m_s}$ a 4x4  spin-isospin
state.


Our points are best illustrated in the simplest $^1S_0$ channel the
extension to other channels being straightforward but cumbersome.  In
the $^1S_0$ channel the scattering process is governed by the
Lippmann-Schwinger equation
\begin{eqnarray}
T_p (k',k) = V (k',k) + \int_0^\Lambda dq q^2 M  
\frac{ V (k',q) T_p (q,k)}{p^2-q^2+i 0^+}  \, , \label{eq:LS} 
\end{eqnarray} 
where $T_p (k',k) $ and $V(k',k)$ are the scattering amplitude and the
potential matrix elements, respectively, between off-shell momentum
states $k$ and $k'$ in that channel.  From the on-shell scattering
amplitude the phase shift, $\delta_0(p)$, can be readily obtained
\begin{eqnarray} 
T_p (p,p) = -\frac{2}{ \pi M p }e^{i \delta_0(p)} \sin \delta_0 (p) \, .
\end{eqnarray} 
The LS Eq.~(\ref{eq:LS}) is solved by standard matrix inversion
techniques. The Momentum space renormalization was addressed in
Ref.~\cite{Entem:2007jg} where we refer for further details.

In coordinate space and for a {\it local} potential $V(r)$ the
on-shell problem can be determined without explicit reference to
off-shell momentum information, although the wave function is obtained
in the non-observable interacting region. The off-shell problem in the
$^1S_0$ channel is formulated as follows from projecting
Eqs.~(\ref{eq:sch-coor}) and (\ref{eq:bc-coor}) onto partial waves.  One
has to solve the Schr\"odinger equation (primes denote derivative with
respect to the radial coordinate $r$) and asymptotic condition 
\begin{eqnarray} -u_p '' (r,k) + U (r) u_p (r,k) &=& p^2 u_p (r,k) 
\nonumber \\ &+& (k^2-p^2) \sin(k r) \, , \\
u_p (r,k) \to \sin (k r) &-& k K(p,k) \cos (pr) \, .  
\label{eq:sch_k} 
\end{eqnarray}
Here $U (r) = 2 \mu_{np} V (r) $ is the reduced potential (in
fact, the Fourier transformation of $V (q)$ ) and $u_p(r,k)$ the
reduced wave function for an s-wave state with energy $p^2/M$ and
momentum $k$. The on-shell case corresponds to take 
\begin{eqnarray}
u_p(r) \equiv u_p
(r,p) \, , \qquad  K(p,p)= \cot \delta_0(p) \, .
\label{eq:on-shell}
\end{eqnarray}
Anticipating the singular character of chiral potentials (see
e.g. Eq.~(\ref{eq:chiral-sing})) these equations are solved for $ r >
r_c $ where $r_c$ is the short distance cutoff which will eventually
be removed, $r_c \to 0$, and the reduced wave function $u_p (r,k)$ is
subject to a suitable boundary condition at $r=r_c$.  What should this
boundary condition be ?. For the finite energy case, $p \neq 0$, it
was argued~\cite{PavonValderrama:2005wv} that completeness of the
on-shell wave functions $u_p(r)$ requires a common domain for the
Hilbert space, requiring
\begin{eqnarray}
\frac{u_p' (r_c)}{u_p (r_c)} &=& \frac{u_0' (r_c)}{u_0 (r_c)} \, , 
\label{eq:bc_sch_on} 
\end{eqnarray}
where the zero energy on-shell problem fulfills 
\begin{eqnarray}
-u_0 '' (r) + U (r) u_0 (r) &=& 0 \, , \qquad r \ge r_c  \, \, \\ 
u_0(r) &\to & 1 - \frac{r}{\alpha_0} \, .
\label{eq:sch_0} 
\end{eqnarray} 
The extended off-shell requirement of a common domain of wave
functions in the physical Hilbert space
\begin{eqnarray}
\frac{u_p' (r_c,k)}{u_p (r_c,k)} &=& \frac{u_0' (r_c)}{u_0 (r_c)} \, , 
\label{eq:bc_sch_off} 
\end{eqnarray}
and correspond to the requirement of completeness of the
on-shell wave functions $u_p(r)$.


We analyze first the simplest case with no potential which in momentum
space corresponds to a contact interaction~\cite{Entem:2007jg}. Since
there is no potential the solutions for $r > r_c$ coincide with the
asymptotic ones, see Eq.~(\ref{eq:sch_k}) and Eq.~(\ref{eq:sch_0}).
Using the relation between the on-shell and off-shell wave
functions at $r=r_c$, Eq.~(\ref{eq:bc_sch_off}), we get
\begin{eqnarray}
K(p,k) = \frac1{k}\, \frac{k (\alpha_0 - r_c) \cos(k
  r_c)+\sin(k r_c)} {\cos(p r_c) + (r_c -\alpha_0 ) p \sin(p r_c)}
\end{eqnarray} 
Note that if $r_c \to 0$ the off-shellness disappears, i.e. we have
$K(p,k) \to K(p,p)$. The momentum space analysis in
Ref.~\cite{Entem:2007jg} yields the same conclusion. This result
suggests that by removing the cut-off we may get rid of the unwanted
off-shell ambiguities.

We turn now to the case of a potential with finite range. Recently,
the finiteness and equivalence of the momentum and coordinate
formulations of the renormalization problem for on-shell
scattering~\cite{Entem:2007jg} and the deuteron bound
state~\cite{Valderrama:2007ja} has been established. Here we extend
those results to the off-shell case. To this end we
follow the insight of previous
works~\cite{PavonValderrama:2004td,PavonValderrama:2007nu} and use the
superposition principle of boundary conditions.  The half off-shell
wave function $u_{p}(r,k)$ can be written as
\begin{eqnarray}
u_p (r,k) = v_p (r,k)  - k K(p,k) w_p (r,k) \, .
\end{eqnarray}
where $v_p(r,k)$ and $w_p(r,k)$ are two auxiliary wave functions 
fulfilling 
\begin{eqnarray}
-v_p '' (r,k) + U (r) v_p (r,k) &=& p^2 v_p (r,k)  \, ,
\nonumber \\ 
v_p(r,k) &\to& \cos (pr) \, , 
\end{eqnarray}
and 
\begin{eqnarray}
-w_p '' (r,k) + U (r) w_p (r,k) &=& p^2 w_p (r,k) \nonumber \\ 
&&+ (k^2-p^2) \sin(k r) \, ,
\nonumber \\ 
w_p (r,k) &\to& \sin (kr) \, , 
\end{eqnarray}
respectively. Our aim is to show that for a singular potential the
function $K(p,k)$ used in Eq.~(\ref{eq:sch_k} ) is finite when the
short distance cut-off is removed, $r_c \to 0 $. We analyze here
the case of a power like {\it attractive} potential, corresponding to
the $^1S_0$ channel considered in the present paper and
represented as $ 2 \mu_{pn} V(r) R^2 \to - (R/r)^n $ where $R$ is a short
distance Van der Waals scale and $n=5,6,7$ corresponds to NLO, N2LO
and N3LO respectively (see Eq.~(\ref{eq:chiral-sing}) and
Ref.~\cite{PavonValderrama:2005wv} for explicit expressions for
$R$). At short distances a WKB approximation
applies~\cite{Case:1950an,Frank:1971xx} and one can show that for $r
\ll R$ one has two independent regular solutions
\begin{eqnarray}
{\cal C}(r) &=& \left(\frac{r}{R}\right)^{n/4} \cos\left[\frac{2}{n-2}
  \left(\frac{R}{r}\right)^{n/2-1} \right] \, ,  \\ 
{\cal S} (r)
&=& \left(\frac{r}{R}\right)^{n/4} \sin\left[\frac{2}{n-2}
  \left(\frac{R}{r}\right)^{n/2-1} \right] \, , 
\end{eqnarray}
fulfilling the Wronskian normalization ${\cal C}'(r){\cal S}(r)-{\cal
  C}(r){\cal S}'(r)= 1/R $.  Note that these asymptotic solutions are
both energy and momentum independent. In terms of these short distance
solutions we must necessarily have at short distances
\begin{eqnarray}
v_p (r,k) &\to & A(p,k)\,  {\cal C} (r) + B(p,k) \, {\cal S} (r) \, , \\ 
w_p (r,k) &\to & C(p,k)\,  {\cal C} (r) + D(p,k) \, {\cal S} (r) \, , \\ 
v_0 (r) &\to & a\,  {\cal C} (r) + b \, {\cal S} (r) \, , \\ 
w_0 (r) &\to & c \, {\cal C} (r) + d\, {\cal S} (r)  \, ,
\end{eqnarray}
where $A(p,k)$, $B(p,k)$,$C(p,k)$ and $D(p,k)$ are suitable energy and
momentum dependent normalization constants, and $a=A(0,0)$,
$b=B(0,0)$, $c=C(0,0)$ and $d=D(0,0)$ the corresponding constants for
the on-shell zero energy problem. From the above equations and the
short distance boundary condition, Eq.~(\ref{eq:bc_sch_off}), it is
straightforward to obtain in the limit $r_c\to 0$ the result
\begin{eqnarray}
k K(p,k) = \frac{\alpha_0 {\cal A}(p,k)+  {\cal B}(p,k)
}{\alpha_0 {\cal C}(p,k)+ {\cal D}(p,k)} 
\end{eqnarray} 
with 
\begin{eqnarray}
{\cal A}(p,k) &=& b A(p,k)  -a B(p,k)  \\ 
{\cal B}(p,k) &=& c B(p,k) - d A(p,k)   \\ 
{\cal C}(p,k) &=& b C(p,k) -a D(p,k)   \\ 
{\cal D}(p,k) &=& c D(p,k)-d C(p,k)     
\end{eqnarray}
which shows explicitly the finiteness of the result. Actually, using
the sub-dominant short distance corrections to the wave functions we
can show that the finite cut-off effect scales as ${\cal O}
(r_c^{n/2-1})$ corresponding for $n=5,6,7$ to a fast convergence. The
previous off-shell relation is a straightforward generalization of the
on-shell result found in Ref.~\cite{PavonValderrama:2005wv}. Similarly
to that case, the off-shell functions ${\cal A}(p,k)$, ${\cal
  B}(p,k)$,${\cal C}(p,k)$ and ${\cal D}(p,k)$ depend by construction
on the potential only. The remarkable feature is the explicit bilinear
dependence on the scattering length, $\alpha_0$. The generalization of
the previous result to higher partial waves and coupled channels is
quite straightforward but cumbersome and will be discussed elsewhere.

\begin{figure}[ttt]
\begin{center}
\epsfig{figure=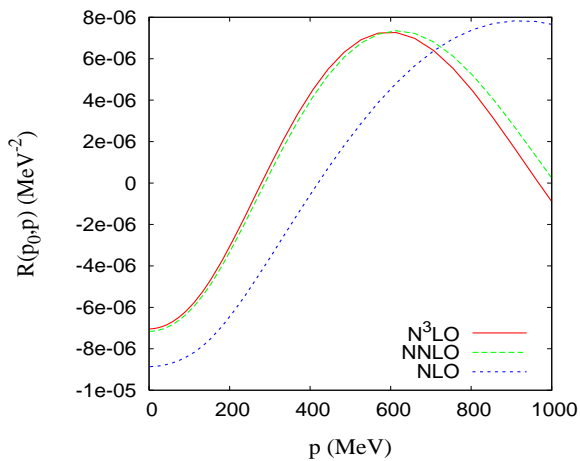,height=6.5cm,width=10cm}
\end{center}
\caption{Renormalized half off-shell R-matrix as a function of the CM
  momentum (in MeV) for $T_{\rm LAB}=50 {\rm MeV}$. The on-shell point
  $p=p_0$ corresponds to $p=153.2 {\rm MeV}$.}
\label{fig:R(p,q)-1S0.eps}
\end{figure}

We turn now to the numerical results. For details on potentials and
parameter choices we refer to Ref.~\cite{Entem:2007jg}. In
Fig.~\ref{fig:R(p,q)-1S0.eps} we show the results for the renormalized
half-off shell $R$-matrix, $R(p,k)=K(p,k)/M$, for a fixed value of the
LAB energy as a function of the CM momentum. We do so for NLO, N2LO
and N3LO, where the potential diverges as $1/r^5$, $1/r^6$ and $1/r^7$
respectively. We have checked the consistency between coordinate and
momentum space NLO and N2LO results provided the same renormalization
conditions are imposed. This confirms the adequacy of requesting the
common boundary condition for both on-shell and off-shell states,
Eq.~(\ref{eq:bc_sch_off}). Besides the sharp cut-off method, we have
also tried a gaussian cut-off, with similar results for off-shell
momenta well below the cut-off range. Despite the strong short
distance singularities our renormalized $R-$matrix does not exhibit
any pathological behaviour and looks as smooth as other
phenomenological and non singular potentials.

%


We summarize our points. We have shown that the renormalizability and
finiteness of the off-shell scattering amplitude for power like short
distance singular potentials rests solely on purely on-shell
information. The out-coming amplitudes are well behaved and soft
despite the underlying short distance singularity being
renormalized. This complies to the desirable expectation that after
renormalization all short distance sensitivity has largely
disappeared, possibly including off-shell ambiguities. Obviously, we
cannot compare our results directly to any experimental quantity as
off-shellness cannot be pinned down by definition. The impossibility
of measuring off-shell effects directly has been emphasized in
Ref.~\cite{Fearing:1999fw} mainly due to the freedom in defining the
physical interpolating field (see also~\cite{Capstick:2007tv}).
While renormalized chiral interactions might be phenomenologically
tested by undertaking three-body, pp-bremsstrahlung or nuclear matter
calculations, it is natural to expect many difficulties. Our 
results suggest a viable and simpler alternative where the
underlying two body singularities are tamed first
through off-shell renormalization and the specific additional
complications of the problem when more than two
bodies are present can be tackled afterwards.


{\it One of us (E.R.A.) thanks Kanzo Nakayama for useful
  communications and L. L. Salcedo for discussions.}



\end{document}